\documentclass[sigconf]{acmart}
\AtBeginDocument{%
  }

\setcopyright{acmlicensed}
\copyrightyear{2025}
\acmYear{2025}
\acmDOI{XXXXXXX.XXXXXXX}
\acmConference[Conference acronym 'XX]{Make sure to enter the correct
  conference title from your rights confirmation email}{June 03--05,
  2025}{Woodstock, NY}
\acmISBN{978-1-4503-XXXX-X/18/06}

\usepackage{multirow}
\usepackage{booktabs}
\usepackage{caption}
\usepackage{tcolorbox}
\usepackage{enumitem}
\usepackage{algorithm}
\usepackage{algpseudocode}

\begin{document}

%%
%% The "title" command has an optional parameter,
%% allowing the author to define a "short title" to be used in page headers.
% \title{RISE: Ranking by Iterative Selection Reduces Position Bias in Large Language Model Recommendations}
\title{Evaluating Position Bias in Large Language Model Recommendations}

%%
%% The "author" command and its associated commands are used to define
%% the authors and their affiliations.
%% Of note is the shared affiliation of the first two authors, and the
%% "authornote" and "authornotemark" commands
%% used to denote shared contribution to the research.
\author{Ethan Bito}
\affiliation{%
  \institution{RMIT University}
  \city{Melbourne}
  \country{Australia}}
\email{s4102812@student.rmit.edu.au}

\author{Yongli Ren}
\affiliation{%
  \institution{RMIT University}
  \city{Melbourne}
  \country{Australia}}
\email{yongli.ren@rmit.edu.au}

\author{Estrid He}
\affiliation{%
  \institution{RMIT University}
  \city{Melbourne}
  \country{Australia}}
\email{estrid.he@rmit.edu.au}

%%
%% By default, the full list of authors will be used in the page
%% headers. Often, this list is too long, and will overlap
%% other information printed in the page headers. This command allows
%% the author to define a more concise list
%% of authors' names for this purpose.
\renewcommand{\shortauthors}{Ethan Bito, Yongli Ren, and Estrid He}

%%
%% The abstract is a short summary of the work to be presented in the
%% article.
\begin{abstract}
Large Language Models (LLMs) are being increasingly explored as general-purpose tools for recommendation tasks, enabling zero-shot and instruction-following capabilities without the need for task-specific training. While the research community is enthusiastically embracing LLMs, there are important caveats to directly adapting them for recommendation tasks. In this paper, we show that LLM-based recommendation models suffer from position bias, where the order of candidate items in a prompt can disproportionately influence the recommendations produced by LLMs. 
%{\color{red}We propose an evaluation framework for systematic quantifying such biases.} 
First, we analyse the position bias of LLM-based recommendations on real-world datasets, where results uncover systemic biases of LLMs with high sensitivity to input orders. Furthermore, we introduce a new prompting strategy to mitigate the position bias of LLM recommendation models called \textbf{R}anking via \textbf{I}terative \textbf{SE}lection (RISE). We compare our proposed method against various baselines on key benchmark datasets. Experiment results show that our method reduces sensitivity to input ordering and improves stability without requiring model fine-tuning or post-processing.
\end{abstract}

%%
%% The code below is generated by the tool at http://dl.acm.org/ccs.cfm.
%% Please copy and paste the code instead of the example below.
%%

%\ccsdesc[300]{Information systems~Probabilistic retrieval models}
%%
%% Keywords. The author(s) should pick words that accurately describe
%% the work being presented. Separate the keywords with commas.
%% A "teaser" image appears between the author and affiliation
%% information and the body of the document, and typically spans the
%% page.

%%
%% This command processes the author and affiliation and title
%% information and builds the first part of the formatted document.
\maketitle

\section{Introduction}
Large Language Models (LLMs) (e.g. ChatGPT and LLaMA ~\cite{brown2020languagemodelsfewshotlearners, touvron2023llamaopenefficientfoundation}) have displayed strong performance on a range of natural language tasks, motivating recent efforts to adapt them for recommendations~\cite{Zhao_2024, Bao_2023, liu2023chatgptgoodrecommenderpreliminary}. LLM-based recommendation systems have emerged as flexible alternatives to traditional approaches like collaborative filtering and neural ranking models~\cite{DBLP:journals/www/WuZQWGSQZZLXC24}. Leveraging strong zero/few-shot capabilities, researchers have increasingly adopted LLMs in conversational and agent-based settings, enabling interactive user experiences via natural language dialogues ~\cite{Lei_2020, 10.1145/3626772.3657844, wang2024recmindlargelanguagemodel}.

Despite these promising results, an increasing number of studies show that the use of LLMs in downstream tasks must be carefully designed and calibrated. Pre-training on large-scale web data introduces internal biases that can emerge in various forms. Zhao et al. ~\cite{zhao2021calibrate} proposes calibrating the LLM before use to mitigate the impact of inherent biases, and Zhang et al. ~\cite{zhang2023chatgpt} highlights how LLMs exhibit prejudiced behaviour toward sensitive attributes when generating recommendations. In addition to internal biases, LLMs are notably sensitive to variations in input prompts. Scaler et al. ~\cite{sclarquantifying} shows that LLMs are sensitive to spurious prompt features that are unrelated to the task. Xu et al. ~\cite{xu2024incontextexampleorderingguided} demonstrates that the order of in-context demonstration examples can lead to drastically different performances of LLMs on downstream tasks.

\begin{figure}[t]
  \centering
  \caption{Illustration of Position Bias in LLM Recommendation. The LLM maps input candidate lists to output rankings. Due to position bias, reversing the input order yields different output rankings for the same items.}
  \includegraphics[width=0.95\linewidth]{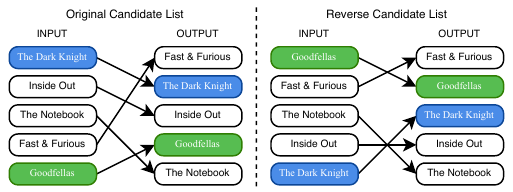} \label{fig:intro}
  \vspace{-0.5cm}
\end{figure}

In this paper, we present a comprehensive study on \textbf{\textit{position bias}} of LLMs, which is particularly relevant to the development of LLM-based recommendation models. We formulate the recommendation task as a learning-to-rank problem, exemplified by an LLM prompt such as: ``Based on the user's preferences, can you rank the following items, \texttt{item1}, \texttt{item2},...?". Such prompt formulation has been widely used to build recommendation models based on LLMs, with promising results. However, in this paper, we show that LLMs are surprisingly sensitive to the order of candidate input items in the above prompt. As illustrated in Fig. ~\ref{fig:intro}, the same input set of candidate items can yield substantially different ranking outputs when the order of input items are reversed. We refer to such sensitiveness of LLMs to input item order as LLMs' \textit{position bias}: the tendency of LLMs to rely on the order of candidate input items rather than their relevance to the prompt. 
 
To the best of our knowledge, there has been no systematic study of position bias in LLMs when applied to recommendation tasks. This paper presents the first in-depth investigation of this phenomenon and highlights the need to calibrate position bias when employing LLMs for recommendation. We focus on the sensitivity of LLMs to prompt variations in the context of ranking tasks, and investigate how the minor changes in the appearance order of candidate items affect the recommendation result produced by LLM recommenders. We show that LLM models are extremely sensitive to the input item orders across various settings, highlighting the presence of strong position bias in LLMs. Furthermore, we propose a simple yet effective prompting technique to alleviate such position bias through \textbf{R}anking via \textbf{I}terative \textbf{SE}lection (RISE). RISE addresses position bias by reducing the original ranking task into smaller, more manageable subtasks, which are then solved in an iterative manner. Our experimental results on real-world datasets show that RISE can effectively reduce the position bias by up-to 25\% compared to baselines. 
% The contributions of this paper are as follows.
% \begin{itemize}
%     \item Systematically evaluation of the position bias of LLM-based recommendations.
%     \item A positional consistency metric to quantify the sensitivity of LLMs to input order.
%     \item The RISE prompting strategy to mitigate position bias and improve ranking consistency.
% \end{itemize}

\section{Related Work}
 Generative LLMs for recommendation refers to the use of natural language to perform recommendation tasks, and can be categorised into two paradigms: non-tuning and tuning-based approaches ~\cite{wu2024surveylargelanguagemodels, Bao_2023, friedman2023leveraginglargelanguagemodels, yang2023palrpersonalizationawarellms}. While tuning-based methods have displayed stronger performance, they often require extensive resources and task-specific data. Non-tuning approaches take advantage of strong LLM zero/few-shot capabilities ~\cite{kojima2023largelanguagemodelszeroshot, brown2020languagemodelsfewshotlearners}, positioning them as lightweight alternatives. Dai et al. \cite{Dai_2023} conducts an analysis of ChatGPT's recommendation ability on three learning-to-rank strategies, and Liu et al. \cite{liu2023chatgptgoodrecommenderpreliminary} systematic evaluates five common recommendation tasks, both proposing prompting frameworks based on their findings. Moreover, Sanner et al. ~\cite{sanner2023largelanguagemodelscompetitive} demonstrates that LLMs can match item-based recommendations in near cold-start scenarios using zero/few-shot prompts on natural language preferences.

Position bias refers to the dependency on the position of items within a candidate list that disproportionately affects a model’s output, often disregarding the relevance of individual items ~\cite{shi2025judgingjudgessystematicstudy}. In LLM-based recommendation tasks, items that appear earlier in the candidate list are more likely to be favoured due to their position. The “Lost in the Middle” phenomenon further highlights similar findings where LLMs return the correct output when answers are located at the beginning and end of documents ~\cite{liu2023lostmiddlelanguagemodels}. Moreover, Wang et al. ~\cite{wang2023largelanguagemodelsfair} shows that the quality of rankings can be gamed by altering the order of where items appear, and Xu et al. ~\cite{xu2024incontextexampleorderingguided} finds that with in-context learning (ICL), the order of examples profoundly impacts the model’s performance. To address position bias in LLM-based recommendations, Hou et al. ~\cite{hou2024largelanguagemodelszeroshot} uses bootstrapping to partially mitigate the bias by aggregating the output of randomly shuffled candidate lists. Ma et al. ~\cite{ma2023largelanguagemodelsstable} provides a two-stage Bayesian framework that uses a probing stage to detect position bias patterns, and a Bayesian adjustment step to calibrate outputs.

% {\color{red}Unlike existing methods that emphasise ranking performance, we further evaluate how prompting strategies affect consistency. We then introduce a mitigation framework that addresses bias without requiring fine-tuning or post-processing.}

% \subsection{Popularity Bias}
% Popularity bias in recommender systems refers to the tendency to favour already highly interacted items at the expense of lesser-known, long-tail items \cite{Klimashevskaia_2024, 10.1145/3109859.3109912}. This bias is often self-reinforced as popular items are repeatedly recommended, accumulating more interactions which further amplifies their visibility and continues pushing less popular content deeper into obscurity \cite{Boratto_2021, 10.1145/3447548.3467376}. Similarly, large language models (LLMs) are also susceptible to popularity bias. Due to their pre-training on large-scale web corpora. LLMs tend to be more familiar with widely referenced and high-frequency content, whilst under-representing lesser-known items \cite{lichtenberg2024largelanguagemodelsrecommender}. We examine popularity bias by evaluating multiple model performances across different segments of item popularity distributions (e.g., top, middle, bottom). This allows us to measure how LLMs respond to variations of item popularity.

\section{Methodology}
\subsection{Problem Formulation}\label{sec:problem_formulation}
We formalise the recommendation problem as a learning-to-rank task. Given a user's interacted items \(I = \{i_1, i_2, \ldots, i_N\}\), the task is to rank the items from the candidate list \(C = \{c_1, c_2, \ldots, c_K\}\) based on the user's preference and generate the ranking list $R=\{r_1, r_2, \ldots, r_K\}$.

We employ pre-trained LLMs as the recommendation model. LLMs are trained to predict the next token $t_s$ given previous tokens $t_1, \ldots, t_{s-1}$ by maximising the likelihood function $P_{\theta}(t_s|t_1,\ldots,t_{s-1})$. Here, $\Theta$ represents the parameters of the LLM. Thus, we prompt the LLM and generate the ranking list $R$ by sampling from:
\begin{equation}
    P_{\theta}(R|\texttt{INST}, I, C) = \prod_{k=1}^K P_{\theta}(r_k|\texttt{INST}, I, C, r_1, \ldots, r_{k-1})
    \label{eq:LLM_sampling}
\end{equation}
with some temperature. Here, \texttt{INST} represents the instruction to the LLM that we append to the prompt. We define this standard prompting template following \textit{list-wise prompting} \cite{Dai_2023, chao2024makelargelanguagemodel, yang2023palrpersonalizationawarellms}. Below illustrates an example template for a movie recommendations system:

\scriptsize
\begin{tcolorbox}[colback=gray!10, colframe=black!30, boxsep=1pt, left=1pt, right=1pt, top=1pt, bottom=1pt, rounded corners]
\textbf{The user has previously watched the following movies:}\\
John Wick, Gone in 60 Seconds, WALL-E, Mad Max: Fury Road, Big Hero 6

\vspace{4pt}
\textbf{Here is a list of candidate movies:}
\begin{itemize}[leftmargin=1em]
  \item[-] The Fast and the Furious
  \item[-] Inside Out
  \item[-] The Dark Knight
  \item[-] The Notebook
  \item[-] Goodfellas
\end{itemize}

\vspace{4pt}
Rank all candidate movies based on the user's preferences.
\end{tcolorbox}
\normalsize

\subsection{Evaluation of Position Bias}
\begin{definition}
    In the context of LLM recommendations, \emph{Position Bias} refers to the influence of the order of candidate items in the input prompt on the resulting rankings generated by a large language model (LLM). Specifically, variations in the order of items in the candidate list $C$ can lead to different output rankings $R$.
\end{definition}
%\textbf{Definition of Position Bias.} We refer to position bias in the context of LLM recommendations as the influence of the order of candidate items in the input prompt on the resulting rankings. That is, changes to the order of items in the candidate list $C$ can lead to different output rankings $R$.   
% \textbf{Definition of Position Bias.} {\color{red}As discussed, LLMs are sensitive to the token order that appears in the prompt. In the context of recommendation task, we define - REMOVE, MAKE CONCISE AND STRAIGHT TO DEF.}position bias as the impact of candidate items' positions in the prompt on the resultant recommendation result. That is, the impact of the item order in $C$ on the generated ranking list $R$.   

\textbf{Quantifying Position Bias.} We quantify the effect of position bias on recommendation outcomes by systematically prompting the LLM with inputs that differ in the positional order of candidate items. Then, we measure the divergence between the resulting recommendations to assess the sensitivity of the LLM model to the input order. 
Specifically, 
given a user $u$, their historical interaction set $I$, and a candidate item list $C$, we prompt the LLM $T$ times. In each iteration, we randomly shuffle the order of items in $C$ and generate a recommendation list using Eq.~\ref{eq:LLM_sampling}. We then compute the pairwise similarity between the $T$ generated lists to quantify the stability of the LLM's recommendations. A high variance across the outputs indicates a strong sensitivity to item order, which reflects the degree of position bias in the model. 
%We summarise the algorithm in Algo.~\ref{alg:evaluation_position_bias}. %{\color{red}- Explanation is already very good, possible remove algorithm 1 depending on space.}

\textbf{Similarity Metric.} In the $T$ generated lists, we compute the similarity between each pair of lists using Kendall's tau coefficient, which measures the rank correlation between two ordered lists:
\begin{equation}\label{eq:kendalls_tau}
    \tau =\frac{n_c - n_d}{\frac{1}{2}n\times (n-1)} 
\end{equation}
where $n_c$ represents the number of concordant pairs (pairs ordered the same way in both rankings) and $n_d$ represents the number of discordant pairs (pairs ordered differently in two rankings). We then aggregate the similarities across all pairs. A low average Kendall's tau indicates high sensitivity to item order. Note that other similarity metrics can also be adapted here. 

% \begin{algorithm}[t]
% \caption{Evaluation of Position Bias}
% \label{alg:evaluation_position_bias}
% \textbf{Input:} 
% Large Language model (LLM) parameterised by $\theta$, Dataset consisting of user-item interaction history $D$
% \hrule
% \begin{algorithmic}[1]
% \For{each author $u$ in $\mathcal{U}$} 
%     \State $I_u \gets \text{obtain interaction history from } \mathcal{D}$
%     \State $C_u \gets \text{construct candidate list}$ 

%     \State $\mathcal{R} \gets \emptyset$
%     \For{$t$ in $T$} \Comment{Repeat $T$ times}
%         \State $C_t \gets \text{shuffle } C_u$
%         \State $R_t \gets \text{Generate ranking list using Eq.~\ref{eq:LLM_sampling}}$ 
%         \State $\mathcal{R} \gets \mathcal{R}\cup \{R_t\}$
%     \EndFor
%     \State Compute pairwise similarity of $\mathcal{R}$ using Eq.~\ref{eq:kendalls_tau}
% \EndFor 
% \end{algorithmic}
% \end{algorithm}

\begin{figure}[tbp]
  \centering  
  \caption{Overview of RISE. 
  %LLM returns one recommendation which is appended to ranked list $R$ and removed from candidate set $C$. The process repeats until all items are ranked.
  \label{fig:rise}}
  \includegraphics[width=0.9\linewidth]{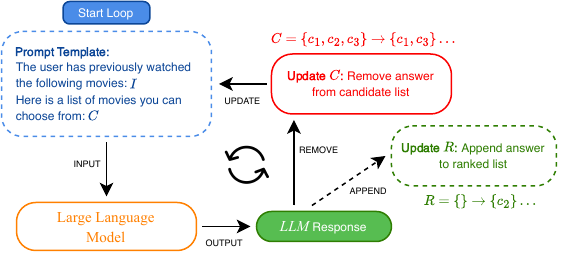}
  \Description{Diagram showing iterative selection using an LLM to incrementally build a ranked list.}
  \vspace{-0.5cm}
\end{figure}

\subsection{RISE: Ranking via Iterative Selection}
The standard prompt template, as described in Section~\ref{sec:problem_formulation}, results in high position bias (see Section~\ref{sec:experiments}). To alleviate the position bias of LLM recommenders, we propose a new prompting technique that generates \textbf{R}anking via \textbf{I}teration \textbf{SE}lection (RISE). The core idea of our prompting technique is to reduce the size of the ranking task and solve it iteratively, following a ``reduce-and-conquer'' strategy. 
Specifically, as shown in Fig.~\ref{fig:rise}, RISE incrementally constructs a ranked list by prompting the model to return a single item at each step. In this way, it simplifies the recommendation task by enabling the model to return one item at a time, and guides the model to reason over the candidate list recursively.

Given a user's interacted items, $I$ and a candidate list $C$, we prompt the LLM to return a single item based on the user's interacted items:
\begin{equation}
    r_k \thicksim P_{\theta}(r|\texttt{INST}, I, C)
    \label{eq:RISE_sampling}
\end{equation}
The returned item $r_k$ is then appended to the ranking list $R$ and removed from candidate list $C$. In the next iteration, the LLM is then prompted to select another item from $P_{\theta}(r|\texttt{INST}, I, C\setminus\{r_k\})$.
We repeat this process until the entire candidate list is ranked. Here, the instruction $\texttt{INST}$ that we append to the prompt is changed to \textit{``Recommend exactly one movie from the candidate list. \ldots"}. 

\textbf{RISE@N.} We further explore how scaling the number of selections impacts recommendation quality by extending the standard iterative selection strategy to allow the model to return $N$ items at each iteration. That is, at each iteration the model selects the most relevant $N$ items from the remaining candidate set. The $N$ items are appended to the ranked list and removed from the candidate pool. We repeat this process until all candidates have been ranked. In our experiments, we evaluate values of \(N \in \{1, 3, 5\}\).
Accordingly, the $\texttt{INST}$ in Eq.~\ref{eq:RISE_sampling} that we append to the prompt for RISE@N is changed to \textit{``Recommend exactly $N$ movie(s) from the candidate list. \ldots"}.

\section{Experiments}
\label{sec:experiments}
\subsection{Experiment Setup}
\label{sec:section_4.1}
\textbf{Datasets.} We run experiments on two popular datasets: (1) \textit{MovieLens-1M}~\cite{10.1145/2827872} which contains 1 million user ratings from 6,000 users on 4,000 movies; and (2) \textit{Amazon Books}~\cite{He_2016}, containing over 22 million user ratings from 8 million users on 2.3 million items.

\textbf{Candidate and User Sampling.} To construct each evaluation sample, we sort items by popularity and divide them into $K$ equal-sized bins. One item is randomly selected from each bin to form a candidate list of size $K$. For each candidate list, we identify a user who has rated at least three of the candidate items. The user's top three rated items within the candidate set serve as ground truth, selected relative to their own ratings rather than an absolute threshold. The user's interaction history is drawn from their highest-rated items outside the candidate list. This sampling process is used across all candidate distributions outlined in Section ~\ref{sec:sub_section_4.4.1}.

\begin{table*}[t]
\centering
\caption{Results on position bias with full candidate distribution. 
%($K$ is the candidate sizes, , and 
(Metrics are reported as mean $\pm$ standard deviation). 
%PC = Positional Consistency, Sim = Output Similarity, Sens = Input Sensitivity.
}
\label{tab:main_results}
\footnotesize
\begin{tabular}{c|c|c|ccccc}
\hline
\textbf{Dataset} & \textbf{Method} & \textbf{K} & \textbf{PC} $\uparrow$ & \textbf{Sim} $\uparrow$ & \textbf{Sens} $\downarrow$ & \textbf{Recall@5} $\uparrow$ & \textbf{NDCG@5} $\uparrow$\\
\hline
\multirow{9}{*}{\centering MovieLens} 
  & \multirow{3}{*}{\centering Standard}   & 10 & 0.67 $\pm$ 0.19 & 0.71 $\pm$ 0.19 & 0.22 $\pm$ 0.17 & 0.72 $\pm$ 0.26 & 0.66 $\pm$ 0.26 \\
  &                                          & 20 & 0.55 $\pm$ 0.18 & 0.59 $\pm$ 0.20 & 0.18 $\pm$ 0.17 & 0.55 $\pm$ 0.30 & 0.50 $\pm$ 0.30 \\
  &                                          & 30 & 0.47 $\pm$ 0.17 & 0.51 $\pm$ 0.19 & 0.16 $\pm$ 0.18 & 0.43 $\pm$ 0.31 & 0.40 $\pm$ 0.30 \\
\cline{2-8}
  & \multirow{3}{*}{\centering Bootstrapping} & 10 & 0.66 $\pm$ 0.20 & 0.79 $\pm$ 0.16 & 0.20 $\pm$ 0.15 & 0.72 $\pm$ 0.26 & 0.67 $\pm$ 0.26 \\
  &                                          & 20 & 0.56 $\pm$ 0.16 & 0.72 $\pm$ 0.13 & 0.14 $\pm$ 0.11 & 0.55 $\pm$ 0.29 & 0.52 $\pm$ 0.29 \\
  &                                          & 30 & 0.48 $\pm$ 0.17 & 0.68 $\pm$ 0.12 & 0.11 $\pm$ 0.09 & 0.45 $\pm$ 0.31 & 0.42 $\pm$ 0.30 \\
\cline{2-8}
  & \multirow{3}{*}{\centering RISE@1}      & 10 & 0.75 $\pm$ 0.10 & 0.77 $\pm$ 0.12 & 0.19 $\pm$ 0.14 & 0.76 $\pm$ 0.23 & 0.71 $\pm$ 0.23 \\
  &                                          & 20 & 0.72 $\pm$ 0.09 & 0.75 $\pm$ 0.09 & 0.13 $\pm$ 0.10 & 0.63 $\pm$ 0.27 & 0.58 $\pm$ 0.27 \\
  &                                          & 30 & 0.69 $\pm$ 0.11 & 0.78 $\pm$ 0.12 & 0.12 $\pm$ 0.09 & 0.50 $\pm$ 0.28 & 0.45 $\pm$ 0.28 \\
\hline
\multirow{9}{*}{\centering Amazon Books} 
  & \multirow{3}{*}{\centering Standard}   & 10 & 0.55 $\pm$ 0.24 & 0.63 $\pm$ 0.20 & 0.27 $\pm$ 0.19 & 0.78 $\pm$ 0.26 & 0.75 $\pm$ 0.27 \\
  &                                          & 20 & 0.48 $\pm$ 0.20 & 0.52 $\pm$ 0.20 & 0.21 $\pm$ 0.18 & 0.62 $\pm$ 0.33 & 0.60 $\pm$ 0.33 \\
  &                                          & 30 & 0.47 $\pm$ 0.16 & 0.49 $\pm$ 0.18 & 0.17 $\pm$ 0.15 & 0.53 $\pm$ 0.33 & 0.52 $\pm$ 0.32 \\
\cline{2-8}
  & \multirow{3}{*}{\centering Bootstrapping} & 10 & 0.56 $\pm$ 0.22 & 0.72 $\pm$ 0.19 & 0.21 $\pm$ 0.15 & 0.78 $\pm$ 0.26 & 0.76 $\pm$ 0.26 \\
  &                                          & 20 & 0.50 $\pm$ 0.19 & 0.67 $\pm$ 0.15 & 0.15 $\pm$ 0.11 & 0.63 $\pm$ 0.32 & 0.61 $\pm$ 0.32 \\
  &                                          & 30 & 0.47 $\pm$ 0.15 & 0.63 $\pm$ 0.12 & 0.12 $\pm$ 0.09 & 0.52 $\pm$ 0.33 & 0.51 $\pm$ 0.32 \\
\cline{2-8}
  & \multirow{3}{*}{\centering RISE@1}      & 10 & 0.65 $\pm$ 0.17 & 0.74 $\pm$ 0.16 & 0.21 $\pm$ 0.16 & 0.72 $\pm$ 0.28 & 0.68 $\pm$ 0.28 \\
  &                                          & 20 & 0.64 $\pm$ 0.11 & 0.72 $\pm$ 0.09 & 0.13 $\pm$ 0.10 & 0.56 $\pm$ 0.35 & 0.53 $\pm$ 0.34 \\
  &                                          & 30 & 0.65 $\pm$ 0.10 & 0.70 $\pm$ 0.10 & 0.11 $\pm$ 0.08 & 0.44 $\pm$ 0.29 & 0.42 $\pm$ 0.28 \\
\hline
\end{tabular}
\normalsize
\end{table*}

\begin{table*}[t]
\centering
\caption{Effect of candidate list distribution on performance (LLaMA 3.3 70B, $K{=}10$). Metrics are reported as mean $\pm$ standard deviation.}
\label{tab:distribution_comparison}
\footnotesize
%\scriptsize
\begin{tabular}{c|ccccc|ccccc}
\hline
\multirow{2}{*}{\textbf{Distribution}}
& \multicolumn{5}{c|}{\textbf{Standard}}
& \multicolumn{5}{c}{\textbf{RISE@1}} \\
\cline{2-11}
& \textbf{PC} $\uparrow$ & \textbf{Sim} $\uparrow$ & \textbf{Sens} $\downarrow$ & \textbf{Recall@5} $\uparrow$ & \textbf{NDCG@5} $\uparrow$ 
& \textbf{PC} $\uparrow$ & \textbf{Sim} $\uparrow$ & \textbf{Sens} $\downarrow$ & \textbf{Recall@5} $\uparrow$ & \textbf{NDCG@5} $\uparrow$ \\
\cline{1-11}
Full        & 0.67 $\pm$ 0.19 & 0.71 $\pm$ 0.19 & 0.22 $\pm$ 0.17 & 0.72 $\pm$ 0.26 & 0.66 $\pm$ 0.26
            & 0.75 $\pm$ 0.10 & 0.77 $\pm$ 0.12 & 0.19 $\pm$ 0.14 & 0.76 $\pm$ 0.23 & 0.71 $\pm$ 0.23 \\
Top         & 0.64 $\pm$ 0.18 & 0.70 $\pm$ 0.20 & 0.24 $\pm$ 0.18 & 0.70 $\pm$ 0.26 & 0.63 $\pm$ 0.27
            & 0.76 $\pm$ 0.10 & 0.78 $\pm$ 0.12 & 0.19 $\pm$ 0.14 & 0.71 $\pm$ 0.26 & 0.65 $\pm$ 0.27 \\
Middle      & 0.66 $\pm$ 0.20 & 0.71 $\pm$ 0.19 & 0.22 $\pm$ 0.17 & 0.76 $\pm$ 0.25 & 0.71 $\pm$ 0.25
            & 0.75 $\pm$ 0.11 & 0.78 $\pm$ 0.12 & 0.19 $\pm$ 0.14 & 0.84 $\pm$ 0.21 & 0.80 $\pm$ 0.20 \\
Bottom      & 0.64 $\pm$ 0.19 & 0.67 $\pm$ 0.22 & 0.23 $\pm$ 0.17 & 0.76 $\pm$ 0.25 & 0.72 $\pm$ 0.25
            & 0.74 $\pm$ 0.11 & 0.77 $\pm$ 0.13 & 0.19 $\pm$ 0.14 & 0.85 $\pm$ 0.20 & 0.82 $\pm$ 0.19 \\
Intertwined & 0.63 $\pm$ 0.23 & 0.90 $\pm$ 0.14 & 0.22 $\pm$ 0.16 & 0.72 $\pm$ 0.25 & 0.65 $\pm$ 0.26
            & 0.74 $\pm$ 0.13 & 0.97 $\pm$ 0.06 & 0.19 $\pm$ 0.15 & 0.78 $\pm$ 0.22 & 0.72 $\pm$ 0.22 \\
\cline{1-11}
\end{tabular}
\normalsize
\end{table*}

\textbf{Baselines.} We compare our proposed prompting technique, RISE, against two key baselines.
$\bullet$ \emph{Standard Prompting:} A simple prompting strategy where the model is asked to rank the entire candidate list in a single pass.
$\bullet$ \emph{Bootstrapping:} A repetition-based extension of standard prompting, designed to reduced position bias by prompting the model $T$ times with randomly shuffled candidate lists ~\cite{hou2024largelanguagemodelszeroshot}. The final ranking is aggregated from multiple iterations. In this paper, we set $T=9$ for bootstrapping, and group the outputs into three sets of three, each of which is aggregated into a final ranking using Borda Count ~\cite{10.1145/371920.372165}.
    % \item \textbf{Bootstrapping:} A repetition-based extension of standard prompting, designed to reduced position bias by prompting the model $T$ times with randomly shuffled candidate lists ~\cite{hou2024largelanguagemodelszeroshot}. The final ranking is aggregated from multiple iterations. {\color{red}In this paper, we set $T$ as 3. We generate nine ranked lists by applying standard prompting over nine shuffled candidate lists. These outputs are grouped into three sets, each containing three ranked lists. For each group, we aggregate the three rankings into a single list using Borda count [x paper?], creating three final aggregated rankings. We report the average results across these three rankings.}

\subsection{Evaluation Metrics}
We evaluate baselines and our proposed approach on several metrics. $\bullet$ \textit{Positional Consistency} (PC): Our positional consistency metric measures the extent to which the model's rankings are influenced by input order, computed using Kendall's tau between outputs generated from original and reversed candidate lists (see Algo. ~\ref{alg:position_bias_metric}). A higher correlation indicates lower position bias and reduced sensitivity to input ordering. $\bullet$ \textit{Output Similarity} (Sim): Measures consistency of model's outputs across multiple runs of shuffled candidate lists using the average pairwise Kendall's tau. $\bullet$ \textit{Input Sensitivity} (Sens): Compute Kendall's tau between the candidate input list and the ranked output list. $\bullet$ \textit{Recall@K} and \textit{NDCG@K} - We adopt standard top-k ranking accuracy metrics that evaluate the presence and ordering of ground truth items in the model's top-k results.

\begin{figure}[tb]
  \centering
  \caption{{
  %\color{red}There is also a 2x2 figure commented out. Might help with space.} 
  Positional Consistency and Output Similarity by prompting strategy. Bars (left to right) show K @ 10, 20, 30.}}
  \includegraphics[width=1.0\linewidth]{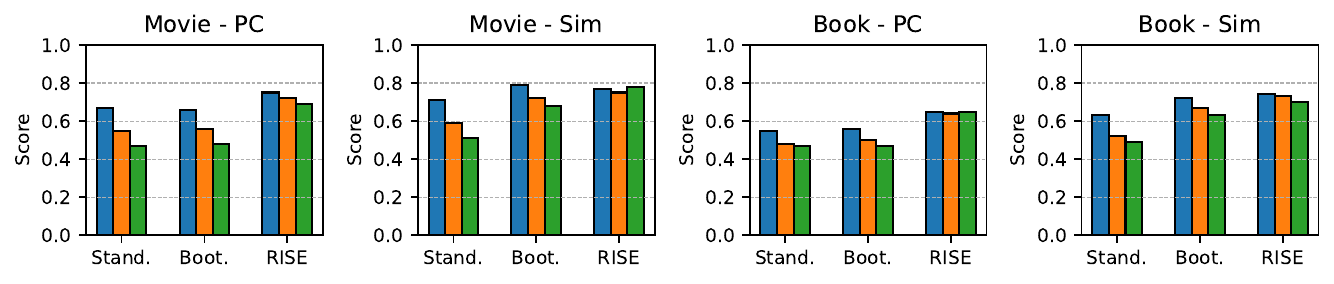}
  \label{fig:pc_sim_comparison}
  \vspace{-0.8cm}
\end{figure}

\footnotesize
\begin{algorithm}[h]
\caption{\footnotesize Positional Consistency Evaluation}
\label{alg:position_bias_metric}
\textbf{Input:} 
Large Language Model (LLM) parameterised by $\theta$, Dataset consisting of user-item interaction history $\mathcal{D}$
\hrule
\begin{algorithmic}[1]
\For{each user $u$ in $\mathcal{U}$}
    \State $\mathcal{I}_u \gets$ obtain interaction history from $\mathcal{D}$
    \State $\mathcal{C}_u \gets$ construct candidate list
    \For{each iteration $i = 1$ to $T$}
        \State $\mathcal{C}_{\text{shuffled}} \gets$ randomly shuffle $\mathcal{C}_u$
        \State $\mathcal{C}_{\text{reversed}} \gets$ reverse($\mathcal{C}_{\text{shuffled}}$)
        \State $R_1 \gets$ generate ranking list from $\mathcal{C}_{\text{shuffled}}$ using Eq. \ref{eq:LLM_sampling}
        \State $R_2 \gets$ generate ranking list from $\mathcal{C}_{\text{reversed}}$ using Eq. \ref{eq:LLM_sampling}
        \State Compute similarity between $R_1$ and $R_2$ using Eq. \ref{eq:kendalls_tau}
    \EndFor
\EndFor
\end{algorithmic}
\end{algorithm}
\normalsize

\subsection{Position Bias Results}
Table ~\ref{tab:main_results} presents the performance of each prompting strategy across three different candidate sizes \(K \in \{10, 20, 30\}\), evaluated using LLaMA 3.3 70B on both MovieLens and Amazon Books datasets. Across these datasets and values of K, iterative selection achieves the highest positional consistency and output similarity. For the MovieLens dataset, iterative selection maintains a positional consistency between 0.75 $\pm$ 0.10 at $K=10$, and 0.69 $\pm$ 0.11 at $K=30$. Similarly in Amazon Books, it ranges from 0.65 $\pm$ 0.17 and 0.64 $\pm$ 0.11. Unlike standard and bootstrapping approaches which display clear degradation in positional consistency and related metrics as K increases, iterative selection remains exceptionally stable, as shown in Fig. ~\ref{fig:pc_sim_comparison}. LLaMA 3.3 70B's performance using standard prompting drops from 0.67 at $K=10$ to 0.47 at $K=30$ on the MovieLens dataset, and 0.55 to 0.57 on the Amazon Books dataset respectively. Bootstrapping follows a similar trend though outperforming standard prompting on most metrics. For accuracy metrics \textit{Recall@5} and \textit{NDCG@5}, standard and bootstrapping exhibit occasionally higher results, particularly in Amazon Books. In the MovieLens dataset however, iterative selection remains competitive and outperforms baseline approaches across all metrics. Thus, iterative selection offers a more favourable trade-off by maintaining competitive ranking quality while greatly reducing position bias and improving similarity consistency. 

\subsection{Position Bias vs Popularity Bias} 

Here, we aim to examine the relationship between position bias and popularity bias, with a particular focus on how popularity bias influences position bias. To conduct a comprehensive investigation, we design five sampling strategies based on different popularity distributions to generate candidate lists: 
$\bullet$ \textit{Full} - samples across the entire popularity distribution. $\bullet$ \textit{Top} - selects from the most popular 20\%. $\bullet$ \textit{Middle} - samples from the 21st to 49th percentiles. $\bullet$ \textit{Bottom} - includes the least popular 50\%. $\bullet$ \textit{Intertwined} - alternates between top and bottom percentiles in the pattern \([0, n-1, 1, n-2, 2, n-3, \ldots]\). 
Then, we evaluate each prompting technique as shown in Table~\ref{tab:distribution_comparison}.

It is observed that variations in positional consistency and output similarity across these distributions are relatively modest. This suggests that popularity bias may not be a primary driver of LLM prompting behaviour in recommendations. The \textit{Middle} and \textit{Bottom} distributions achieve highest accuracy metrics, indicating slightly better ranking performance on less popular items. However, the \textit{Intertwined} distribution results in the highest output similarity score. We note that unlike other distributions, the intertwined candidate lists are not shuffled in the prompts in order to preserve their alternating structure. This lack of variation leads to artificially high similarity scores as the model is exposed to identical input sequences. Overall, while input distribution does influence output stability to some extent, these effects are relatively modest and do not suggest a strong or consistent preference towards item popularity in the current setup.

% It is observed that variations in positional consistency and output similarity across these distributions are relatively modest. This suggests that popularity bias may not be a primary driver of LLM prompting behaviour in recommendations. The top distribution consistently returns the lowest metrics compared to other sampled distributions, where candidate items in this range are strongly recognisable, potentially making the models rankings sensitive to the input order. This is further validated by the top distribution also having highest input sensitivity of 0.21 $\pm$ 0.15, suggesting difficulty in ranking movies due to their popularity. In contrast, the intertwined distribution results in the highest output similarity score. We note that unlike other distributions, the intertwined candidate lists are not shuffled in the prompts in order to preserve their alternating structure. This lack of variation leads to artificially high similarity scores as the model is exposed to identical input sequences. Overall, while input distribution does influence output stability to some extent, these effects are relatively modest and do not suggest a strong or consistent preference towards item popularity in the current setup.
\label{sec:sub_section_4.4.1}

\subsection{RISE@N Evaluation} We examine the effects of iterative selection by $N$ on performance in Table ~\ref{tab:iterative_by_n}. We see clear degradation in positional consistency and ranking quality as the value of $N$ increases. RISE@1 consistently produces the greatest results compared to other values of $N$, confirming that more precise step-wise selection is most effective.

\begin{table}[tb]
\scriptsize
\centering
\caption{Effect of iterative selection depth ($N$) ($K=20$, Full Distribution). Metrics are reported as mean $\pm$ standard deviation.}
\label{tab:iterative_by_n}
\scriptsize
\begin{tabular}{c|ccccc}
\hline
\textbf{RISE@$N$} & \textbf{PC} $\uparrow$ & \textbf{Sim} $\uparrow$ & \textbf{Sens} $\downarrow$ & \textbf{Recall@5} $\uparrow$ & \textbf{NDCG@5} $\uparrow$\\
\hline
1 & 0.72 $\pm$ 0.09 & 0.75 $\pm$ 0.09 & 0.13 $\pm$ 0.10 & 0.63 $\pm$ 0.27 & 0.58 $\pm$ 0.27 \\
3 & 0.68 $\pm$ 0.08 & 0.72 $\pm$ 0.08 & 0.14 $\pm$ 0.10 & 0.59 $\pm$ 0.27 & 0.55 $\pm$ 0.27 \\
5 & 0.61 $\pm$ 0.09 & 0.67 $\pm$ 0.10 & 0.16 $\pm$ 0.11 & 0.60 $\pm$ 0.28 & 0.54 $\pm$ 0.28 \\
\hline
\end{tabular}
%\vspace{-0.2cm}
\normalsize
\end{table}

\section{Conclusion}
Our study further demonstrates how LLM-based recommenders are sensitive to the input order of candidate items. We propose an iterative selection mitigation strategy that incrementally constructs ranked lists, aiding in reducing the effects of position bias accentuated by LLMs. Results illustrate that iterative selection consistently outperforms baselines prompting approaches on key bias, consistency, and accuracy metrics. 
%{\color{red}Furthermore, our approach focuses on mitigating position bias without the use fo fine-tuning or post-processing, addressing the bias throughout each prompt iteration. We show how a simplistic yet effective alleviation strategy can meaningfully influencing the fairness and reliability of LLM recommendations.} 

% \section*{GenAI Usage Disclosure}

% This research complies with the CIKM 2025 GenAI usage policy. The authors disclose the following use of Generative AI (GenAI) tools during the research process:

% \begin{itemize}
%   \item \textbf{Writing:} ChatGPT (GPT-4, OpenAI) was used to assist in proofreading, rephrasing technical sentences, and improving the clarity of the manuscript. All substantive content, including ideas, methods, results, and analysis, was written and verified by the authors.
%   \item \textbf{Code:} No GenAI tools were used to generate any code in this research. All code was developed by the authors.
%   \item \textbf{Data:} No GenAI tools were used to generate or augment the data used in this research. All datasets were obtained from publicly available sources as described in the paper.
%   \item \textbf{Experiments and Analysis:} No GenAI tools were used to generate experimental results or statistical analyses.
% \end{itemize}

% The authors confirm that all intellectual contributions are original and that the use of GenAI tools did not compromise the scientific integrity or originality of the work.

%%
%% The next two lines define the bibliography style to be used, and
%% the bibliography file.
\bibliographystyle{ACM-Reference-Format}
\bibliography{sample-base}

%%
%% If your work has an appendix, this is the place to put it.

\end{document}